\documentclass[final,5p,times,twocolumn]{elsarticle}

\usepackage{amssymb}
\usepackage{amsmath}
\usepackage{listings}
\usepackage{verbatim}
\usepackage{url}
\usepackage{eurosym}
\usepackage{bm}
\usepackage{bbm}

\usepackage[colorlinks=true, pdfstartview=FitV, linkcolor=blue, citecolor=blue, urlcolor=blue, breaklinks=true]{hyperref}

\journal{Computer Physics Communications}

\newcommand{\changed}[1]{\textcolor{black}{#1}}
\newcommand{\added}[1]{\textcolor{black}{#1}}

\newcommand{\ba}{{\bm a}}
\newcommand{\bU}{{\bm U}}
\newcommand{\bF}{{\bm F}}

\begin{document}

\begin{frontmatter}



\title{Performance of a high-order MPI-Kokkos accelerated fluid solver}


\author[label1,label2]{Filipp Sporykhin}
\author[label1]{Holger Homann\corref{author}} 

\affiliation[label1]{organization={Université Côte d’Azur, Observatoire de la Côte d’Azur, CNRS, Laboratoire Lagrange},
            addressline={Boulevard de l’Observatoire CS 34229 - F 06304 NICE Cedex 4, France}}
\affiliation[label2]{organization={MAUCA - Master track in Astrophysics, Université Côte d’Azur \& Observatoire de la Côte d’Azur},
            addressline={Parc Valrose, 06100 Nice, France}}

\begin{abstract}
  This work discusses the performance of a modern numerical scheme for
  fluid dynamical problems on modern high-performance computing (HPC)
  architectures. Our code implements a spatial nodal discontinuous
  Galerkin (NDG) scheme that we test up to an order of convergence of
  eight. It is temporally coupled to a set of Runge-Kutta (RK) methods
  of orders up to six. \changed{The code integrates the linear
    advection equations as well as the isothermal Euler equations in
    one, two, and three dimensions.} In order to target modern
  hardware involving many-core Central Processing Units (CPUs) and
  accelerators such as Graphic Processing Units (GPUs) we use the
  Kokkos library in conjunction with the Message Passing Interface
  (MPI) to run our single source code on various NVidia and AMD GPU
  systems.
  
  By means of one- and two-dimensional simulations of simple test
  equations we find that the higher the order the faster is the
  code. Eighth-order simulations attain a given global error with much
  less computing time than third- or fourth-order simulations. The RK
  scheme has a smaller impact on the code performance and a classical
  fourth-order scheme seems to generally be a good choice.

  The code performs very well on all considered HPC GPUs. We observe
  very good scaling properties up to 64 AMD MI250x GPUs and \added{we
    show that the scaling properties are the same in two and three
    dimensions.}  The many-CPU performance is also very good and
  perfect weak scaling is observed up to many hundreds of CPU cores
  using MPI. We note that small grid-size simulations are faster on
  CPUs than on GPUs while GPUs win significantly over CPUs for
  simulations involving more than $10^7$ degrees of freedom ($\approx
  3100^2$ grid points). When it comes to the environmental impact of
  numerical simulations we estimate that GPUs consume less energy than
  CPUs for large grid-size simulations but more energy on small
  grids. Further, we observe a tendency that the more modern is the
  GPU the larger needs to be the grid in order to use it
  efficiently. This yields a rebound effect because larger simulations
  need longer computing times and in turn more energy that is not
  compensated by the energy efficiency gain of the newer GPUs.
\end{abstract}




\end{frontmatter}


\lstset{language=C++,frame=single}
\lstset{breaklines=true,basicstyle=\footnotesize}
\section{Introduction}

Numerical simulations of fluid flows are today an indispensable tool
for solving scientific and engineering problems. Often these flows
exhibit great complexity as is for example the case for turbulent
flows~\cite{frisch1995turbulence, Pope_2000}. Turbulence is a very
active domain of research and affects many parts of the physical
world, from cloud physics~\cite{shaw-2003}, the formation of
planetesimals in protoplanetary disks~\cite{birnstiel-2024}, the
effects of Earth's atmosphere on optical wavefronts
\citep{atmosphere}, to the interactions between turbulent solar winds
and bodies in the solar system (\cite{mars_turbulence},
\cite{mercury_turbulence} and \cite{solar_turb_magneto}) or
engineering problems such as the design of
wind-turbines~\cite{neuhaus-2024}. Simulations of turbulent flows
require very large numerical resolutions in order to resolve all flow
structures from the large to the small scales. This in turn, leads to
numerical degrees of freedom counts (typically grid points or cells)
in the billions, and long computation times. Fluid simulations can be
very expensive. This difficulty can only be overcome with efforts from
both the software and the hardware side. Software means that numerical
schemes for solving fluid equations need to be improved and hardware
means that these better schemes run efficiently on faster, modern
computer architectures.

Let us start with the hardware perspective.  In the past,
supercomputers have typically assembled many Central Processing Units
(CPUs) (consisting each of many computing units called cores). The
early 21st century has then seen a massive increase in consumer grade
Graphics Processing Units (GPUs), originally designed to perform very
specific shader and floating point calculations in graphical
rendering. Today, GPUs are used as accelerators for demanding
scientific computations. The main difference between the two lies in
the larger internal cache and higher clock speed of CPUs, and the
larger core count of GPUs (typically $10^4$ computing units). GPUs
excel at doing the same instruction on a large data set asynchronously
\citep{owens2008gpu}. GPUs are currently replacing CPUs in
high-performance supercomputers \cite{top500} due to their higher
floating point operation (FLOP) performance. \textit{A priori} they
are also more power efficient, which can be seen by looking at the top
10 today's supercomputers on the Green500 (\cite{green500}) list,
which all highly make use of GPUs. In this work, we will investigate
the energy efficiency of CPUs and GPUs for the present numerical
scheme.

A drawback of GPUs is that they require in principle vendor specific
written code. NVidia develops the language \emph{CUDA} for its GPUs
while AMD's language is called \emph{HIP}. CUDA or HIP written code
does not run on CPUs. Codes developed in CUDA require some effort to
port when moving to different hardware. When hardware changes one may
have to spend a significant amount of time to adapt code to new
architecture. To tackle this issue, there have recently been
\ attempts to develop a unifying language that allows to write a
single source code that runs on CPUs as well as GPUs from different
vendors, namely OpenMP~\cite{openmp}, RAJA~\cite{raja,raja_basics},
\added{OpenACC~\cite{openacc}} and the library we will be using in
this work: Kokkos~\citep{kokkos,kokkos_basics}. \added{All these
  languages have been successfully applied in HPC
  computations~\cite{gross-geosx,XU2017577,XU2023123649,kokkos-qcd,davis2024evaluative}.}

Now, let us turn to the software part. The efficiency of different
numerical schemes that solve the same fluid equations vary at lot. An
indicator of the efficiency is the order of convergence of the
numerical scheme~\cite{kreiss1972comparison}. Generally, the higher
the order the faster is the scheme and the smaller are the memory
requirements (see section~\ref{subsec:serial}). Among the different
numerical schemes that can reach high orders of convergence we have
chosen a nodal Discontinuous Galerkin (NDG) type scheme. It is well
suited for integrating conservation laws and is local in the sense
that data dependencies are minimal. This feature is well adapted to
modern supercomputers for which network bandwidth limitations can
penalize data exchanges resulting in bottlenecks.

In essence NDG is similar to the Finite Elements (FE) method
\citep{zienkiewicz-etal2014}, but it is discontinuous across cells and
employs the numerical flux commonly found in Finite Volumes (FV)
schemes \citep{LeVeque_2002}. It considerably differs from FV in the
way how it achieves high orders by avoiding the reconstruction
process.

This paper is organized as follows. We first present in
section~\ref{sec:methos} the used methods by starting with the Kokkos
library (section~\ref{subsec:kokkos}) followed by the nodal
discontinuous Galerkin scheme (section~\ref{sec:ndg}) and the
Runge-Kutta methods (section~\ref{sec:rk}). We then turn in
section~\ref{sec:results} to the discussion of the obtained
results. Here, we start with the serial properties of the numerical
RK-NDG scheme (section~\ref{subsec:serial}) that is it's properties on
single-core machines. After that it's performance on highly parallel
architectures is discussed in section \ref{subsec:parallel}. We finish
the results section with environmental considerations in
section~\ref{subsec:env}. Conclusions and perspectives are drawn in
section~\ref{sec:conc}.

\section{Methods}
\label{sec:methos}

\subsection{The nodal discontinuous Galerkin scheme}
\label{sec:ndg}

For the spatial discretization of the fluid equations we have chosen a
nodal discontinuous Galerkin scheme (NDG)~\cite{hesthaven-warburton}
for mainly three reasons: First, it is natively high-order and
therefore very efficient. Second, it is local so that it is expected
to scale well on massive parallel GPU accelerated
supercomputers. Third, NDG is conservative and therefore well suited
for conservation laws arising in fluid dynamics.


Let us illustrate the main ideas of NDG schemes with a 1D scalar
conservation law on the physical domain $\Omega=\{x \in [0,L]\}$:

\begin{equation}
\label{eq:scalar_cons}
    \frac{\partial U}{\partial t} + \frac{\partial F(U)}{\partial x} = 0
\end{equation}
Here, $U = U(x,t)$ is the conserved variable with initial profile
$U(x,t=0)=U_0(x)$ and $F(U)$ the flux function. More generally, one
could encounter a source term on the right hand side of
(\ref{eq:scalar_cons}) that we ignore here for simplicity. When
discretizing, we partition the global domain $\Omega$ into $N$
elements $\Omega_j$ such as $\Omega=\cup^{N}_j\Omega_j$ with $\Omega_j
= [x_{j-1/2},x_{j+1/2}]$, $j=1,...,N$.



The main idea of discontinuous Galerkin schemes (the signification of
'nodal' will be introduced a little bit later) is to approximate the
solution $U_j(x,t)$ in a cell $\Omega_j$ by a polynomial of degree $p$

\begin{equation}
\label{eq:proj_poly}
    U_j(x,t) = \sum_{k=0}^p c_j^k(t) h_k(x)
\end{equation}

\begin{figure}
  \centering
  \includegraphics[width=0.48\textwidth]{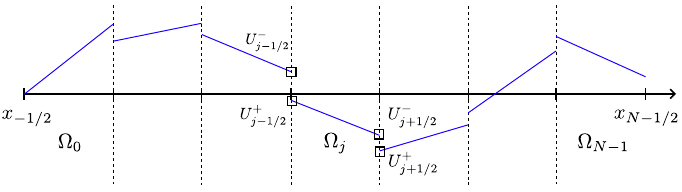}
  \caption{Sketch of a linear (first-order) polynomial representation
    of a sinusoidal $U(x,t)$ profile. Discontinuities can be seen
    between cells. The $-$ and $+$ signs denote the flux values the
    left and right polynomial at the intersection of two cells.}
  \label{fig:discretisation_with_poly}
\end{figure}

Here, the polynomial coefficients $c_j^k$ are the discrete unknowns of
every cell and $h_k(x)$ are polynomials of order $k$. As an example of
a discretized signal we show a sinusoidal initial $U_o$ in
Fig.~\ref{fig:discretisation_with_poly} using a linear polynomial
($p=1$). In this case, $c_0$ and $c_1$ (corresponding to cell-average
and slope) are the two unknowns per cell. For comparison, in
finite-volume schemes the only unknown is the average cell value $\bar
U_j=\frac{1}{x_{j+1/2}-x_{j-1/2}}\int_{\Omega_j} U(x) dx$. Using only
cell averages leads to a spatially first-order scheme. In
finite-volume methods, higher order schemes are obtained by
reconstruction of higher-order polynomials in each cell by using cell
averages of adjacent cells. This is omitted in NDG schemes as the
higher-order polynomial coefficients are part of the set of unknowns.

We get an equation for the NDG unknowns $c_j^k$ by projecting the
equation (\ref{eq:scalar_cons}) onto a polynomial $h_l$, which is
called the weak formulation. Restricting for a moment our attention to
the first term of (\ref{eq:scalar_cons}) the projection leads to
\begin{equation}
  \label{eq:firstTerm}
  \sum_{k=0}^p\int_{\Omega_j} \dot c_j^k(t) h_k(x) h_l(x) dx.
\end{equation}
In order to isolate the coefficients $c_j^k(t)$ it is convenient to
choose an orthogonal set $h_k$ of polynomials that obeys
\begin{equation}
\label{eq:lag_coefs}
\int_{\Omega_j} h_k(x) h_l(x) dx \approx \sum_{l=0}^{p} h_k(x_l)
h_l(x_l) w_l \approx w_k \delta_{kl},
\end{equation}
where the integral has been evaluated by a $p+1$-point quadrature with
quadrature points $x_l$ and weights $w_l$.  The first term
(\ref{eq:firstTerm}) now simply reads
\begin{equation*}
  \dot c_j^k(t) w_k.
\end{equation*}
Note that there is no sum over $k$ because of the orthogonal
projection.

Before treating the second term of (\ref{eq:scalar_cons}) we come to
the notion of \emph{nodal} discontinuous Galerkin methods. For those,
the set of $h_k$ is chosen to be the Lagrange polynomials

\begin{equation}
  \label{eq:lagrange_pol}
    h_k(\xi) = \frac{(\xi-1)(\xi+1)P_N'(\xi)}{N(N+1)P_N(\xi_k)(\xi-\xi_k)}
\end{equation}

on the interval $\xi \in [-1, 1]$ with nodes $\xi_k$ at the
Gauss-Lobatto quadrature points that are generated by solving
$(1-\xi^2)P_N'(\xi)=0$. $P_N(\xi)$ is the Legendre polynomial of order
$N$. The Lagrange polynomials display the following property at the
nodes:

\begin{equation}
\label{eq:ortho_at_roots}
    h_k(\xi_l)=\delta_{kl} = \left\{\begin{array}{l}
      1 ~\text{if} ~k=l\\
      0 ~\text{if} ~k \neq l
    \end{array}\right.
\end{equation}

An important feature for the performance of the numerical scheme is
that the Gauss-Lobatto quadrature points (for example $[-1, 0, 1]$ for
a quadratic polynomial) include the cell borders meaning that we are
able to directly access the border-values of $U$ in order to compute
the numerical flux that will be discussed in the following paragraph.
Note that the interval coordinates $\xi$ and $x$ are related by the
simple transformation

\begin{equation}
\label{eq:transformation}
x=\frac{\Delta x}{2}\xi+\frac{\Delta x}{2}, \quad dx = \frac{\Delta
  x}{2} d\xi.
\end{equation}

Now we continue with the second term of
(\ref{eq:scalar_cons}). Projection and partial integration yields
\begin{equation*}
  \int_{\Omega_j} \frac{\partial F(U_j)}{\partial x} h_l(x) dx=
  -\int_{\Omega_j} F(U_j)\frac{\partial h_l(x)}{\partial x} dx +
  [F(U_j)h_l(x)]_{x_{j-1/2}}^{x_{j+1/2}}.
\end{equation*}
Combining the first and second term yields the evolution equation for
the unknown coefficients $c_j^k(t)$
\begin{equation}
  \dot c_j^k(t) w_k -\int_{\Omega_j}  F(U_j)\frac{\partial h_l(x)}{\partial x} dx + [F(U_j)h_l(x)]_{x_{j-1/2}}^{x_{j+1/2}} = 0.
\end{equation}
The integral term in the above equation is usually computed by
Gauss-Lobatto quadrature while the third term contains the flux
$F(U_j)|_{x_{j-1/2}}^{x_{j+1/2}}$ at the cell borders $x_{j\pm
  1/2}$. As $U_j$ is discontinuous at the cell borders the flux is
discontinuous, too, and requires to be modeled by the so-called
\emph{numerical flux} $\hat{F}$ obeying $F(U_j)|_{x_{j\pm
    1/2}}\approx\hat{F}(U_{j\pm 1/2}^-,U_{j\pm 1/2}^+)$, $U_{j\pm
  1/2}^-$ and $U_{j\pm 1/2}^+$ being the left and right values of $U$
at either of the cell borders $x_{j\pm 1/2}$ (see
Fig.~\ref{fig:discretisation_with_poly}). The exact solution to
$\hat{F}$ can be found by solving the so-called \emph{Riemann
problem}~\cite{toro2009riemann}. For our purpose it is sufficient to
use the simple Lax-Friedrichs (LF) numerical flux, that is often used
in NDG schemes~\citep{shu-dg:2009}, which is less accurate than the
Riemann solution but does not reduce the order of the scheme,
\begin{equation}
\begin{split}
\label{eq:LF}
    \hat{F}(U_j^-,U_j^+) &= \frac{1}{2} \left[F(U_j^-) + F(U_j^+)) -
      \alpha_{max}(U_j^+ - U_j^-)\right],
\end{split}
\end{equation}

where $\alpha_{max}$ absolute value of the largest eigenvalue of the
flux Jacobian $F'(U)$ evaluated at the cell border. For our
one-dimensional example this reduces to
\begin{equation*}
  \alpha_{max} = \left|\frac{dF(U)}{dU} \right|.
\end{equation*}

\subsection{The Kokkos library}
\label{subsec:kokkos}

Kokkos~\citep{kokkos} is a C++ abstraction layer in the form of a
template library that is used to translate universal code to different
languages such as standard C++, CUDA, HIP, OpenMP, SYCL, and others
which are adapted to different devices such as CPUs and GPUs. In this
context, the target device is called \emph{backend}. The idea is that
one writes one single source code and Kokkos translates it for
different architectures. Therefore, Kokkos requires \textit{a priori} no code
duplication per device which saves development time. In theory no
further debugging on particular devices (CPUs, NVidia GPUs, AMD GPUs,
Intel GPUs, ...) is required, which is an advantage in the every
changing world of high performance computing. However, as we will see
in section~\ref{subsec:parallel} in practice hardware specific tuning
is sometimes necessary to obtain optimal performance.

In essence, Kokkos uses macros that are translated to code adapted to
a specific device.  The particular code can, for example, be a CUDA
kernel. Let us briefly compare the syntax of standard C++ code and
Kokkos code for a very simple, one-dimensional example that could be
seen as a \emph{position based} initialization of a vector containing
\verb+n+ elements. Let us start with a standard C++
\verb+for+-loop approach:

\begin{lstlisting}[caption={\label{lst:1dcpp} C++ \emph{for-loop} initialization.}]
  std::vector<double> vec(n);
  for(int i=0; i<n; i++)
    vec[i] = i;
\end{lstlisting}

Here, the position is represented by the index \verb+i+ so that the
values of the data-holding container \verb+vec+ depends on the
position \verb+i+. The assignment in the above listing is done in an
ordered manner guaranteed by the \verb+i+-incrementation in the
\verb+for+ loop. However, on GPUs operations on many different indices
\verb+i+ are executed preferentially in an unordered and simultaneous
manner. So that this standard C++ code cannot directly be used on a
GPU. Instead, the following Kokkos code runs both on CPUs and GPUs
(even from different vendors):

\begin{lstlisting}[caption={\label{lst:1dkokkos}  \emph{Kokkos} kernel initialization.}]
  Kokkos::View<double*> vec("vec-name", n);
  Kokkos::parallel_for("kernel-name", n,
  KOKKOS_LAMBDA(int i) { vec(i) = i; });
\end{lstlisting}

\verb+Kokkos::View+ is a multi-dimensional container (in the above
example only one-dimensional) holding the data on the computing device
(CPU or GPU). It's constructor takes a name (for profiling and
debugging improvements) and the container size \verb+n+. 
\added{For additional View dimensions that can be added to the
  template argument we must pass the extent of the dimension 
  either via the constructor or via the} \verb+Kokkos::resize+
\added{function.  As can be surmised from the * syntax of the type
  declaration, Views are used as pointers to device data. The
  advantage of this pointer-based approach is the innate compatibility
  with MPI.}

The \verb+parallel_for+ function replaces the \verb+for+ loop in the
standard C++ codelet. This Kokkos codelet could give the impression
that Kokkos code is very different from standard C++ code as in Kokkos
there is no notion of \verb+for+- loops. However, we can rewrite the
codelet \ref{lst:1dcpp} using modern C++23 in order to make it
resemble the Kokkos codelet:
\begin{lstlisting}[caption={\label{lst:1cpp23} Modern \emph{C++23 lambda} initialization.}]
  std::vector<double> vec(n);
  std::ranges::for_each(std::views::iota(0,n-1),
  [&vec](int i){ vec[i] = i;});
\end{lstlisting}
where a standard C++ lambda function performs the element-wise
initialization. The portability of Kokkos now means that for example
in the Kokkos listing \ref{lst:1dkokkos} the macro
\verb+KOKKOS_LAMBDA+ would translate to \verb+__host__ __device__ [=]+ when
compiled for a CUDA device and to a regular C++ lambda capture clause
\verb+[&vec]+ (see listing \ref{lst:1cpp23}) when compiled for the
CPU.

Let us shortly mention on the data layout of the code. We remind that
a one-dimensional nodal discontinuous Galerkin scheme stores the
polynomial coefficients $c_j^k$ (see (\ref{eq:proj_poly})) at $p+1$
nodal points of each cell, where $p$ is the order of the scheme. Let
us consider a two-dimensional situation. If we call $n_x$ the number
of cells in $x-$direction and $n_y$ in $y-$direction and $n_{var}$ the
number of integrated variables ($n_{var}=1$ for linear advection) we
end up with $n_x\times n_y\times p \times p \times n_{var}$ unknowns
for describing the state of the linear advection system. The code
implements this data structure as a \verb+Kokkos::View<double*****>+
with five dimensions. If one deals with several dimensions then the
\verb+Kokkos::parallel_for+ function in listing \ref{lst:1dkokkos}
takes a so-called \verb+Kokkos::MDRangePolicy+ in order to specify the
extent of the data array. All simulations discussed in this work use
double precision floating point data. Additional details on Kokkos can
be found in \cite{kokkos_basics}.

\subsection{Used Runge-Kutta schemes}
\label{sec:rk}

Following \cite{cockburn2001runge} we use explicit Runge-Kutta (RK)
schemes for time integration. One of the goals of this work is to
investigate the influence of the order of the temporal part on the
performance of the overall scheme (discussed in the results section
\ref{subsec:serial}). RK schemes exist for different orders of
convergence and are easy to implement.  Here, we list the three
different Runge-Kutta schemes that we mainly used throughout this
work.

The lowest order scheme is a third-order Runge-Kutta scheme requiring
three sub-steps:

\begin{equation}
  \label{eq:RK3}
    \begin{split}
      k_1 &= dt\,F(U^N)/3 \\
      k_2 &= 2\,dt\,F(U^N + k_1)/3 \\
      k_3 &= dt\,(k_1 + 3\,F(U^N+ k_2))/4 \\
      U^{N+1} &= U^N + k_3
    \end{split}
\end{equation}

We also use the classical fourth-order scheme requiring four
sub-steps:

\begin{equation}
  \label{eq:RK4}
    \begin{split}
      k_1 &= dt\,F(U^N) \\
      k_2 &= dt\,F(U^N + k_1/2) \\
      k_3 &= dt\,F(U^N + k_2/2) \\
      k_4 &= dt\,F(U^N + k_3) \\
      U^{N+1} &= U^N + (k_1 + 2k_2 + 2k_3 + k_4)/6
    \end{split}
\end{equation}

And finally, the highest-order Runge-Kutta scheme~\cite{luther:1968}
that we use is a sixth-order scheme requiring seven sub-steps:
\begin{equation}
\label{eq:RK6}
    \begin{split}
      k_1 &= dt\,F(U^N) \\
      k_2 &= dt\,F(U^N+k_1) \\ 
      k_3 &= dt\,F(U^N+(3k_1+k_2)/8) \\
      k_4 &= dt\,F(U^N+(8k_1+2k_2+8k_3)/27) \\
      k_5 &= dt\,F(U^N + (3(3\sqrt{21}-7)k_1-8(7-\sqrt{21})k_2 \\
      &+48(7-\sqrt{21})k_3 - 3(21-\sqrt{21})k_4)/392) \\
      k_6 &= dt\,F(U^N + (-5(231+51\sqrt{21})k_1 -40(7+\sqrt{21})k_2 \\
      &-320\sqrt{21}k_3 + 3(21+121\sqrt{21})k_4 \\
      &+ 392(6+\sqrt{21})k_5)/1960)\\
      k_7 &= dt\,F(U^N + (15(22+7\sqrt{21})k_1 + 120k_2 \\
      &+ 40(7\sqrt{21}-5)k_3- 63(3\sqrt{21}-2)k_4\\
      &-14(49+9\sqrt{21})k_5+70(7-\sqrt{21})k_6)/180)\\
      U^{N+1} &= U^N+ (9k_1+64k_3+49k_5+49k_6+9k_7)/180
    \end{split}
\end{equation}

\section{Results}
\label{sec:results}
\subsection{The serial RK-NDG scheme}
\label{subsec:serial}
In this section we present the performance properties of the RK-NDG
scheme for serial runs, that is simulations using only one CPU
core. Its parallel performance will be analyzed in the next
section. We start our investigation with the global convergence
behavior of the one-dimensional linear advection equation
\begin{equation}
  \label{eq:advection1d}
  \frac{\partial U}{\partial t} + a \frac{\partial U}{\partial x} = 0,
\end{equation}
$a$ being a constant. It has the form of the one-dimensional
conservation law (\ref{eq:scalar_cons}) with $F(U) =a\,U$. This
equation is often used in order to test numerical schemes in the realm
of fluid dynamics~\cite{toro2009riemann,LeVeque_2002}. It serves
especially as a simple prototype for transport dominated problems
because it's solution is known to be $U_0(x - at)$: The initial signal
$U_0(x)=U(x,t=0)$ is simply transported with the so-called advection
speed $a$.

\begin{figure}[t]
\centering
\includegraphics[width=0.48\textwidth]{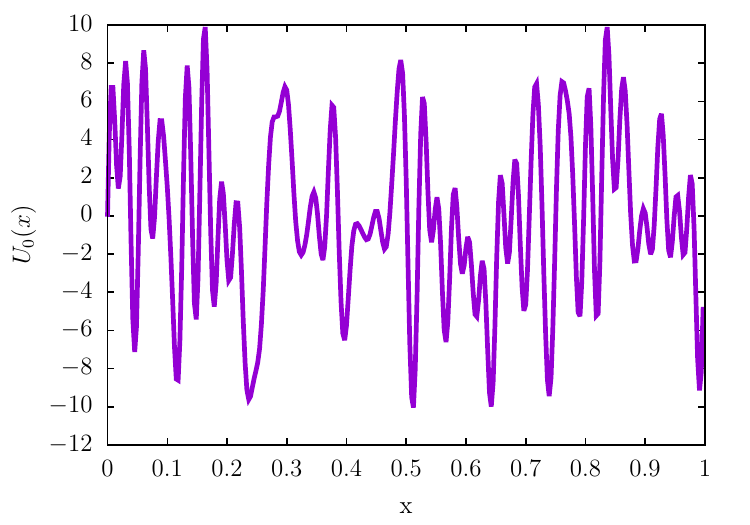}
\caption{Initial condition in the case of
  $N_k=40$.}\label{fig:initalCond_k40}
\end{figure}

Here we set $a=1$ and integrate the equation (\ref{eq:advection1d})
for a time lapse of $T=1$ in order to simulate precisely one
revolution of the initial signal $U_0$. For the latter we choose
\begin{equation}
  \label{eq:initialCond}
  U_0(x) = \sum_{k=1}^{N_k}A_k\sin(2\pi k x),
\end{equation}
$A_k \in [0,1]$ being random amplitudes. The maximal wave number $N_k$
is meant to model the complexity of the considered flow. An image of a
typical initial condition for $N_k=40$ is shown in
Fig.~\ref{fig:initalCond_k40}. \added{We use periodic boundary
  conditions for all simulations: $U^-_{-1/2}=U^-_{N-1/2}$ et
  $U^+_{N-1/2}=U^+_{-1/2}$ with the notation used in
  Fig.~\ref{fig:discretisation_with_poly}.}

For error estimation we use the $L_2$ norm
\begin{equation*}
  L_2 = \sqrt{\int_\Omega \left[U(x, t=1) - U_0(x) \right]^2 dx}
\end{equation*}
$\Omega$ being the computational domain.

\begin{figure}[t]
\centering
\includegraphics[width=0.48\textwidth]{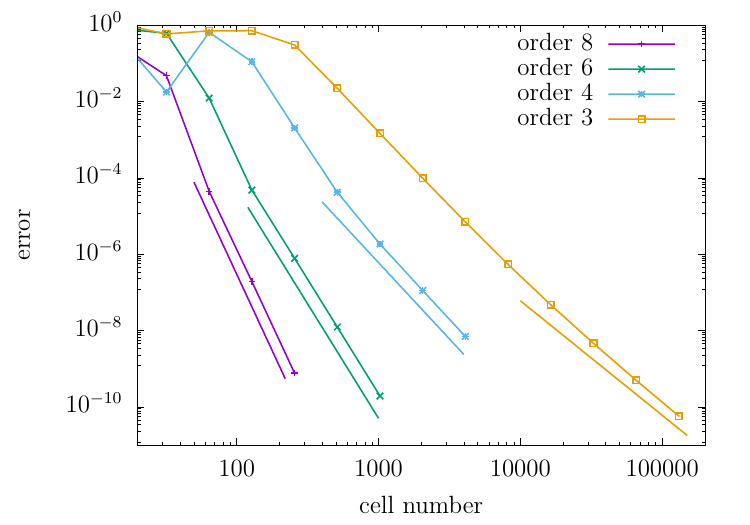}
\caption{Convergence tests for the 1d advection equation with an
  initial condition using $N_k=40$ for different orders. All
  simulations use a sixth-order Runge-Kutta scheme. The straight solid
  lines indicate the expected large cell number
  scaling.}\label{fig:convergence}
\end{figure}

\begin{figure}[h]
\centering
\includegraphics[width=0.48\textwidth]{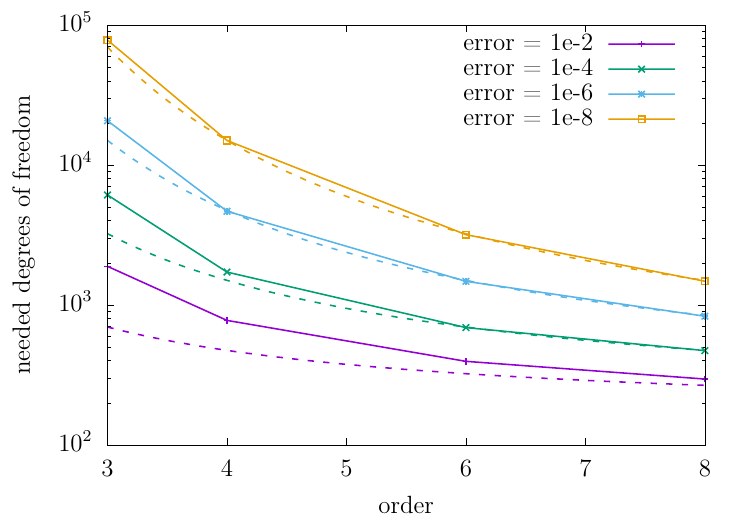}
\caption{Required number of degrees of freedom to achieve a given
  error as a function of the spatial order of the scheme for 1d
  advection for $N_k=40$. The dashed functions are of the form
  $c\,(1/error)^{1/p}$ from Kreiss-Oliger, where the constant $c=200$
  is the same for all graphs}\label{fig:cell_error}
\end{figure}

We observe the expected convergence of the numerical scheme as shown
in Fig.~\ref{fig:convergence}. It is interesting to estimate the
necessary spatial resolution to obtain a given error. How many cells
does one need in order to limit the error to, lets say, the order of
one percent after one revolution? This question is directly related to
the memory requirements of a simulation. The answer is shown in
Fig.~\ref{fig:cell_error}. In order to better compare schemes of
different orders we will consider the number of degrees of freedom
(dof). As we are dealing with a one-dimensional problem in this
section we define (dof) = (number of cells)*(order). One observes in
Fig.~\ref{fig:cell_error} that the higher the order the less degrees
of freedom are needed to obtain a given error. Depending on the
tolerated error the memory requirements of a third-order simulation
are a couple of times up to more than one hundred times bigger than of
an eight-order simulation.

These observations are in agreement with a theoretical estimate of
Kreiss and Oliger~\cite{kreiss1972comparison}. Indeed, the theoretical
fit shown as dashed lines in Fig.~\ref{fig:cell_error} agrees well for
small errors while substantial differences are observed for small
errors. This might be due to the fact that the theoretically assumed
convergence only sets in beyond errors of the order of $10^{-4}$.

The observed superiority of the high-order schemes with respect to
memory usage is even more pronounced in more than one dimensions as in
$d$ dimensions the required degrees of freedom are raised to a power
of $d$. For example, eight-order 2d simulations need roughly 44 times
less degrees of freedom than third-order simulations for one percent
errors. For smaller errors of the order of $10^{-4}$ this ratio
attains 1600.

\begin{figure}[t]
\centering
\includegraphics[width=0.48\textwidth]{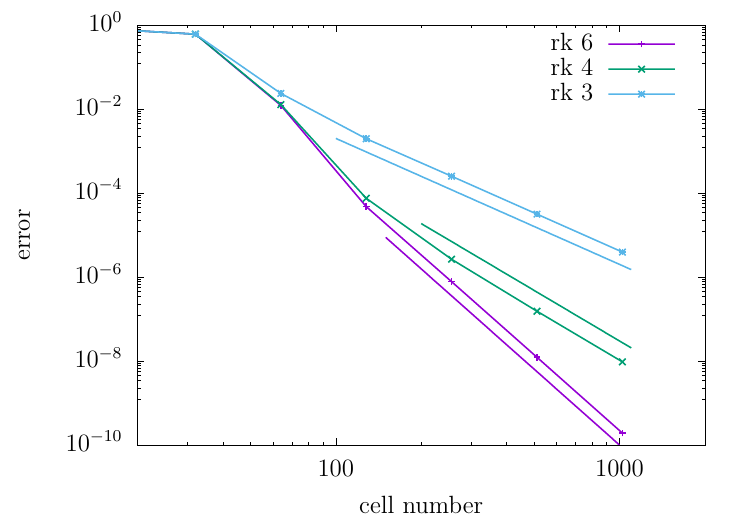}
\caption{Convergence tests for the 1d advection equation with an
  initial condition using $N_k=40$. The spatial order of the scheme is
  six for all runs and the order of the temporal Runge-Kutta scheme is
  varied.}\label{fig:convergence_d_rk}
\end{figure}
Now, we turn to the influence of the order of the temporal scheme. We
show in Fig.~\ref{fig:convergence_d_rk} the convergence properties of
the scheme for a spatially six-order simulation when varying the order
of the used Runge-Kutta scheme. When increasing the spatial resolution
the temporal error starts dominating from a certain resolution on and
reduces the global order of convergence to the order of the employed
Runge-Kutta scheme.

This work is specifically concerned with the performance of high-order
NDG-RK scheme. \changed{In this section we consider two- and three-
  dimensional flows} because performance becomes a concern typically
for dimensions higher than one. \added{We will consider two different
  equations in conservation form that can written as}
\begin{equation*}
  \frac{\partial \bU}{\partial t} + \partial_x \bF_x(\bU) + \partial_y
  \bF_y(\bU) + \partial_z \bF_z(\bU) = 0,
\end{equation*}
\added{where $\bU$ is a vector with $n_{var}$ unknowns and
  $\bF_x(\bU)$, $F_y(\bU)$, $F_z(\bU)$ the fluxes in $x$, $y$ and $z$ direction,
  respectively. The first considered equation, allowing for error
  estimates of the numerical solution is the linear advection for the
  scalar $U$ ($n_{var}=1$) given by the fluxes}
\begin{equation}
  \label{eq:advection2d}
  F_x(U)=a_x U, \quad F_y(U)=a_y U, \quad F_z(U)=a_z U,
\end{equation}
\added{$\ba=(a_x,a_y, a_z)$ being the advection velocity.  The second,
  considered for testing the non-linear performance of the numerical
  scheme are isothermal Euler equations given by} 
\begin{equation}
  \begin{aligned}
  \label{eq:eulerIso2d}
  \bU&=\begin{pmatrix}
  \rho \\
  \rho u_x \\
  \rho u_y \\
  \rho u_z 
  \end{pmatrix}, \quad
  \bF_x=\begin{pmatrix}
  \rho u_x\\
  \rho u_x^2 + \rho a^2 \\
  \rho u_xu_y\\
  \rho u_xu_z 
  \end{pmatrix} \\
  \bF_y&=\begin{pmatrix}
  \rho u_y\\
  \rho u_yu_x \\
  \rho u_y^2 + \rho a^2 \\
  \rho u_yu_z 
  \end{pmatrix}, \quad
  \bF_z=\begin{pmatrix}
  \rho u_z\\
  \rho u_zu_x \\
  \rho u_zu_y \\
  \rho u_z^2 + \rho a^2 
  \end{pmatrix}, 
  \end{aligned}
\end{equation}
\added{$\bU$ thus containing four unknowns ($n_{var}=4$), three
  in the two dimensional case and $a$
  being the sound speed. We will consider subsonic flows such that the
  Mach number is $M=a/u_{max}\approx 0.5$, $u_{max}$ being the maximal
  velocity of the flow. Mostly, we will present results on
  two-dimensional flows (solving only for the $x$ and $y$ components
  of the equations) because the three-dimensional version shows the
  same performance as the two-dimensional as will be shown at the end
  of this section.}

\begin{figure}[t]
\centering
\includegraphics[width=0.48\textwidth]{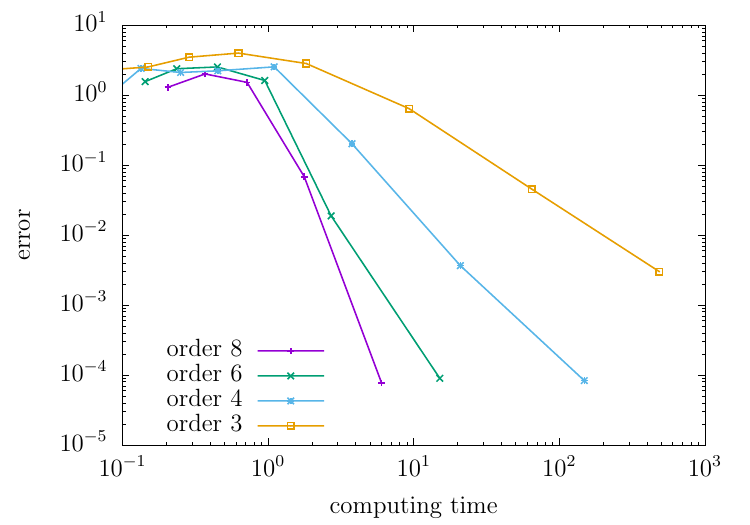}
\caption{Error as a function of the computing time for simulations of
  the two-dimensional linear advection equation (with $N_k=40$) for
  varying spatial order and sixth-order Runge-Kutta scheme. (Run on a
  NVidia V100 GPU)}\label{fig:error_chrono_k40_d}
\end{figure}

\added{We first turn to the order-dependent performance of the
  scheme. For this,} we present in Fig.~\ref{fig:error_chrono_k40_d}
the numerical error \added{from the two-dimensional linear advection
  equation after one-revolution} not depending on the number of grid
points but on the computing time. \added{Similar to the
  one-dimensional case we set $a_x=1, a_y=0$ and integrate the
  equation (\ref{eq:advection2d}) for a time lapse of $T=1$ of the
  initial signal $U_0$. For the latter we choose
\begin{equation}
  \label{eq:initialCond}
  U_0(x,y) = \sum_{k=1}^{N_k}A_k\sin(2\pi k x),
\end{equation}
$A_k \in [0,1]$ being random amplitudes.}

Computing time and error are
relevant parameters to judge the efficiency of a numerical scheme
because the solution of a physical problem typically tolerates a
certain error and the efficiency of a scheme might be expressed as the
time it takes to obtain this solution by means of numerical
simulations.

Although high-order schemes are numerically more expensive per time
step they win over low-order schemes by their increased convergence
rate. Indeed, in order to achieve the same error (one percent for
example) simulating one revolution with the third-order scheme needs
much more time (roughly one hundred times more) than the eight-order
scheme (see Fig.~\ref{fig:error_chrono_k40_d}). The smaller is the
desired error the larger is this ratio. We note that it gets also
larger (not shown) for more complex initial conditions (by increasing
$N_k$ in (\ref{eq:initialCond})).

\begin{figure}[t]
\centering
\includegraphics[width=0.48\textwidth]{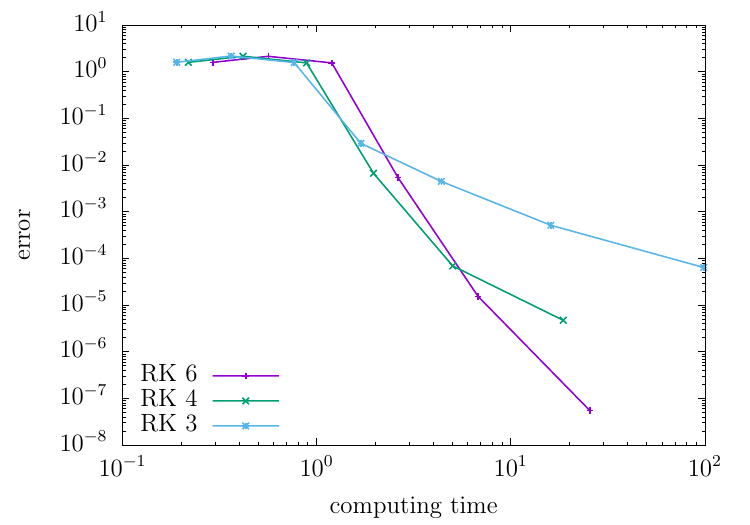}
\caption{Error as a function of the computing time for simulations of
  the two-dimensional linear advection equation (with $N_k=40$) for
  different Runge-Kutta schemes. All simulations are spatially of
  eighth order. (Run on a NVidia A100 GPU)
  GPU.}\label{fig:error_chrono_o8_k40_d}
\end{figure}

In order to estimate the performance implications of the temporal
convergence we show in Fig.~\ref{fig:error_chrono_o8_k40_d} the error
as a function of the computing time for different Runge-Kutta
schemes. For large errors, the third-order RK scheme is the fastest
because it requires less sub-steps (only three) to perform one time
step. But when reducing the error it is quickly overcome by the
classical fourth-order scheme which uses four sub-steps. At very small
errors, the sixth-order Runge-Kutta scheme performs the best although it
uses seven sub-steps. Down to errors of the order of $10^{-5}$ the
differences are quite small so that using the classical fourth-order
Runge-Kutta scheme seems to be a good candidate for many problems.

\subsection{Performance of the parallel RK-NDG scheme on different architectures}
\label{subsec:parallel}
We will now analyze in detail the parallel performance of the RK-NDG
scheme. We will address the efficiency of the scheme when using many
CPU cores or many GPUs.

Starting with CPUs, we performed benchmarks with three different CPU
models (see Tab.~\ref{table:cpu}) that were available on the two
super-computers that we used. The number of cores per CPU ranges from
20 to 96 for the available CPUs. By trend, this number increases with
every next generation.

\begin{table}[t]
  \footnotesize
\centering
\begin{tabular}{c | c c c}
                        & Xeon 6248   & Xeon 8468   & AMD EPYC 9654 \\ 
  \hline                                                         
  physical cores        & 20          & 48          & 96            \\
  SIMD FP64 width       & 8           & 8           & 8             \\
  clock speed           & 2.5 GHz     & 2.1 GHz     & 2.4 GHz       \\
  L3 cache              & 27.5 MByte  & 105 MByte   & 384 MByte     \\
  FP64 performance      & 0.51 TFlops & 2.15 TFlops & 3.9 TFlops    \\
  Thermal Design Power  & 150 W       & 350 W       & 320 W         \\
  Launch year           & 2019        & 2023        & 2022          \\
  Launch price          & 3000 \euro  & 7000 \euro  & 11000 \euro      
\end{tabular}
\caption{Hardware specifications of used CPUs relevant to the present
  work.  'SIMD FP64 width' means the SIMD width for 64 bit floating
  point operations. 'FP64 performance' gives the performance in terms
  of the number of floating point operations per second (Flops) for
  double (64 bit) precision data. The 'Thermal Design Power' serves as
  an estimate for the maximal power consumption of the CPU announced
  by the manufacturer.  The launch prices are only vague estimates as
  they vary rapidly. }\label{table:cpu}
\end{table}

We start the parallel performance considerations with benchmarking the
code on a CPU. For inter-core and inter-CPU communications we use the
well established message passing interface (MPI). It allows to
transfer data between multiple CPUs that do not share the same main
memory (RAM).

\begin{figure}[t]
\centering
\includegraphics[width=0.48\textwidth]{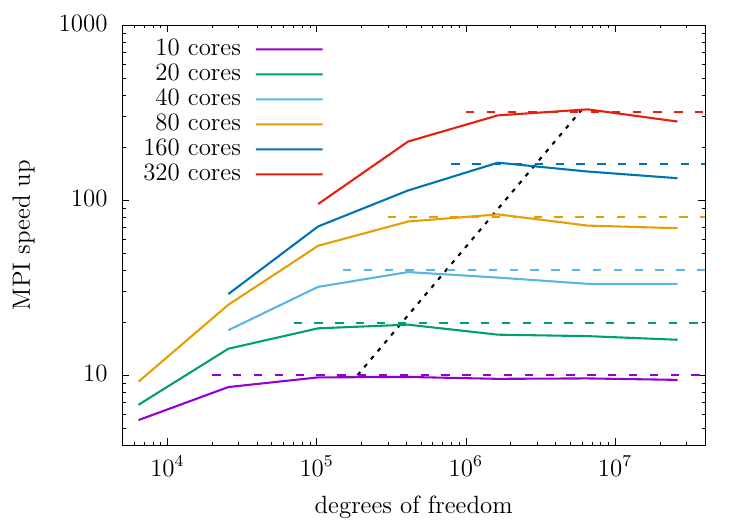}
\caption{Ratio of the computing time of a serial code to the MPI code
  (speed-up) per degree of freedom of the two-dimensional linear
  advection equation. We use an eight-order NDG scheme with sixth-order
  RK. The number of cores used by the MPI runs varies from ten to
  320. One CPU (Xeon 6248) contains 20 cores and one node consists of
  two CPUs so that 320 MPI processes use 8 computing nodes. The values
  on the dashed line exhibit optimal weak scaling of the
  code.}\label{fig:speed_up_mpi}
\end{figure}

To make use of MPI, we subdivide the global domain into subsets and
assign each subset to a different CPU core. \added{The sizes of these
  rectangular (2d) or cuboid (3d) subsets are chosen to comply with
  load balancing.} The communication (for exchanging the border
fluxes) uses non-blocking send/receive operations from the MPI
library. We perform 100 time steps of the two dimensional advection
equation (\ref{eq:advection2d}) on a single core or on many cores
simultaneously and compare the simulation times. The performance gain
(speed-up) from using many cores compared to a single core is shown in
Fig.~\ref{fig:speed_up_mpi} as a function of the number of degrees of
freedom. Apart from small grids, for which the communication overhead
between different cores/nodes deteriorate the performance of many-core
simulations, the code shows very good scaling properties. This is true
up to 320 cores which is the highest number of cores that we
tested. For a certain range of grid-sizes we even observe
\emph{optimal weak-scaling}. In this case the speed-ups are
proportional to the number of cores when keeping the work load per
core constant. Let us give two examples: On only 20 cores (one entire
CPU) the performance is optimal if the simulation uses $\approx
4\times 10^5$ degrees of freedom which amounts to $\approx
640\times640$ grid points for our two-dimensional simulation. For 320
cores the performance is optimal for $\approx6\times 10^6$ degrees of
freedom which corresponds to a $\approx 2500\times 2500$ grid
points. It is important to note that the order of the scheme has no
significant influence on the performance shown in
Fig.~\ref{fig:speed_up_mpi} as long as the number of degrees of
freedom is kept constant. Curves for simulations of different orders
fall on top of each other once they are plotted as a function of the
number of dof. We will therefore often use the number of dof in order
to characterize the size of the simulation.

\begin{figure}[t]
\centering
\includegraphics[width=0.48\textwidth]{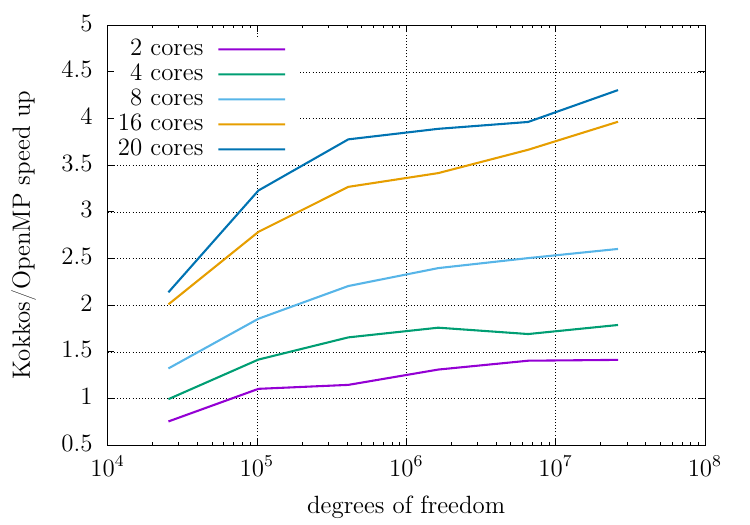}
\caption{Ratio of the computing time of a simulation with the serial
  backend and the OpenMP backend (speed-up) per degree of freedom of
  the two-dimensional linear advection equation. The number of cores
  used by OpenMP runs varies from two to the maximal number (20) of
  available cores on the Xeon 6248 CPU. }\label{fig:time_openMP}
\end{figure}

For the MPI benchmarks we compiled the code for the \emph{serial}
Kokkos backend. Kokkos also comes with an OpenMP backend meaning that
Kokkos code is \textit{a priori} ready for simulations on \emph{shared
memory} machines that is a collection of CPUs sharing the same main
memory. However, OpenMP does not handle data exchange on distributed
memory architectures as MPI does. In Fig.~\ref{fig:time_openMP} we
show the speed-up of the Kokkos OpenMP code with respect to a serial
code running on a single CPU core. We observe that the speed-up is
increasing with the number of degrees of freedom. That is, the larger
is the grid the better performs the multi-core code. However, we only
observe a speed-up of approximately four while for the used CPU (see
Tab.~\ref{table:cpu}) we expect a speed-up of 20 as we observe for the
MPI runs.

We suspect different reasons for this poor OpenMP performance. One
major reason lies in the data layout of the code. We remind here that
\added{in an example of a two dimensional simulation} we store the data with a five-dimensional
\verb+Kokkos::View<double*****>+ view of \changed{geometry $n_x\times
  n_y\times p \times p \times n_{var}$. As $n_x$ and $n_y$ are large
  numbers and $p$ and $n_{var}$ are small numbers } the dimensions of
the data array are imbalanced. This seems to cause problems for
optimal core usage by the OpenMP backend in combination with a flat
\verb+Kokkos::MDRangePolicy+ execution policy. Experiments with
\verb+Kokkos:TeamPolicy+ or other data layouts showed much better
speed-ups that were comparable to the MPI benchmarks. However, these
experiments performed worse on GPUs (to be discussed next). As the
focus of this paper is to compare a single code base on different
architectures we do not go here into more detail on kernel- and
architectural-wise tuning of the code.

\begin{table*}[t]
  \footnotesize
\centering
\begin{tabular}{c | c c c c c c}
                        & NVidia V100 SXM2  & NVidia A100 SXM 4 & NVidia H100 SXM 5 & NVidia RTX A4000  & AMD MI250X   & AMD MI300A   \\ 
  \hline                                                                                                                              
  cores                 & 5120              & 6912              & 16896             &  6144             & 14080        & 14592        \\
  clock speed           & 1.230 GHz         & 0.765 GHz         & 1.59 GHz          &  0.735 GHz        & 1.7 Ghz      & 2.1 Ghz      \\
  FP64 performance      & 7 TFlop           & 9.7 TFlop         & 33.4 TFlop        & 9.6 TFlop         & 47.9 TFlop   & 61.3 TFlop   \\
  Memory                & 16 GByte          & 40 GByte          & 80 GByte          & 16 GByte          & 128 GByte    & 128 GByte    \\
  Thermal Design Power  & 300 W             & 500 W             & 700 W             & 140 W             & 500 W        & 550 W        \\
  Launch year           & 2018              & 2020              & 2023              & 2021              & 2021         & 2024         \\
  Launch price          & 9000 \euro        & 15000 \euro       & 30000 \euro       & 900 \euro         & 15000 \euro  & 20000 \euro  
\end{tabular}
\caption{Hardware specifications of used GPUs relevant to the present
  work. 'FP64 performance' gives the performance in terms of the
  number of floating point operations per second (Flops) for double
  (64 bit) precision data. The 'Thermal Design Power' serves as an
  estimate for the maximal power consumption of the GPU announced by
  the manufacturer. The launch prices are only vague estimates as they
  vary rapidly. }\label{table:gpu}
\end{table*}

We now turn to the performance of the RK-NDG scheme on GPUs. We
analyzed three different generations of HPC NVidia GPUs, one high-end
consumer grade graphics card and two different AMD GPUs (see
Tab.~\ref{table:gpu}). We included the consumer grade graphics card
in order to investigate the performance of a device that costs at
least 10 times less than HPC hardware.

\begin{figure}[t]
\centering
\includegraphics[width=0.48\textwidth]{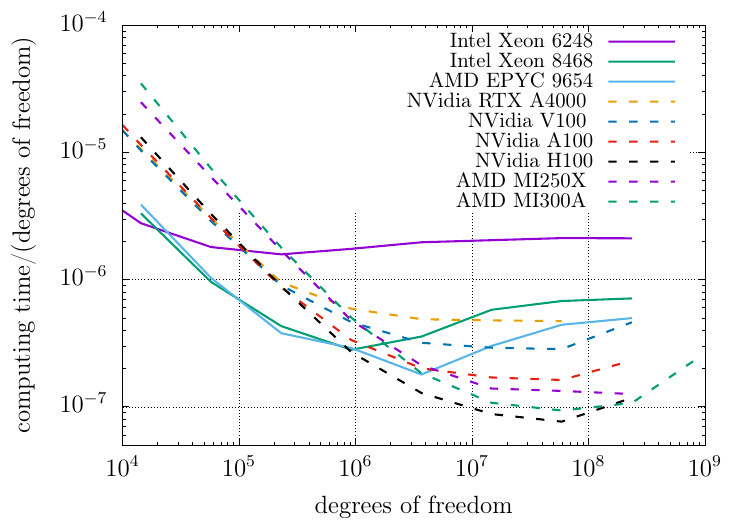}
\caption{Computing time for performing 100 time steps of the
  two-dimensional linear advection equation (\ref{eq:advection2d})
  normalized to the number of degrees of freedom for several CPUs and
  GPUs. }\label{fig:time_cell_d}
\end{figure}

As for CPUs we are interested in the performance as a function of the
number of degrees of freedom. Fig.~\ref{fig:time_cell_d} shows the
computing time needed to perform 100 time steps of the two-dimensional
linear advection equation (\ref{eq:advection2d}) on different
GPUs. For comparison we also include the computing time on the three
different CPUs from table~\ref{table:cpu} when using all of their
physical cores. GPUs are in general faster than CPUs for large grids
while they are slower for small grid. The positions of the cross-overs
depend both on the performance of the CPU and that of the GPU. While
some GPUs outperform the twenty-core CPUs at $1.5\times\,10^5$ degrees
of freedom, the 96 core CPU is competitive up to $4\times\,10^6$
degrees of freedom. We observe also that the more recent the GPU
architecture the more degrees of freedom are needed to take full
advantage of the computing power. While the consumer-grade GPU works
efficiently at $10^6$ degrees of freedom, the latest HPC-GPUs of
NVidia and AMD need several $10^7$ degrees of freedom to attain their
maximal performance. This is probably due to the fact that they employ
more compute units than the consumer grade GPU and therefore need more
data in order to occupy all compute resources. In turn, for the
investigated CPU and GPU configurations, simulations up to roughly
$10^6$ dof can be done efficiently and faster on the CPU. This
corresponds to grids with roughly $400\times 400$ points. Large
simulations with more than $10^7$ degrees of freedom are well adapted
to the HPC GPUs. This corresponds to grids with $3200 \times 3200$
grid points.

In order to further clarify the relative performance of CPUs and GPUs
as a function of the grid-size we now investigate the benefit of one
architecture over the other. This is particularly important if one
speculates on parallelizing a serial code. \changed{Should one go for
  a CPU- or GPU-based implementation?}

\begin{figure}[t]
\centering
\includegraphics[width=0.48\textwidth]{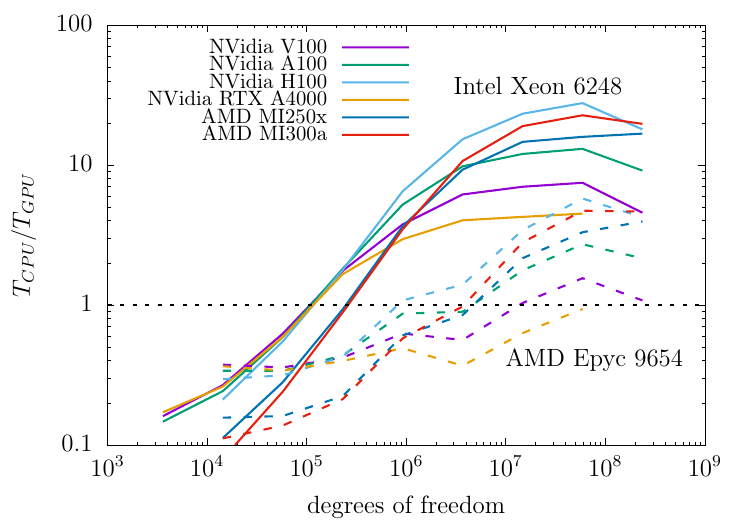}
\caption{Computing time on CPUs compared to GPUs (speed-up). Presented
  is the ratio for two different CPUs (Xeon 6248 with 20 MPI processes
  \& AMD EPYC 9654 with 96 MPI processes) and six different GPUs as a
  function of the number of degrees of
  freedom.}\label{fig:time_gpu-mpi-cell_d}
\end{figure}

From Fig.~\ref{fig:time_gpu-mpi-cell_d} we can see that the speed-up
from using GPUs can be large when compared to a low core number
CPU. When comparing a modern GPU to a less modern CPU, the GPU can do
large simulations about 20-30 times faster. But if one has a high core
number CPU at hand, even for large grid-sizes the GPU performance gain
is quit limited and attains a maximum factor of six. From this one
might question if it is worth to write GPU ready code using Kokkos. It
might be sufficient to parallelize using a standard C++ CPU code. In
this context it is worth noting that the prices of the CPUs and GPUs
are very different. As we deal mostly with HPC hardware, exact prices
are not easily available but rough estimates indicate huge price
differences to the point that the consumer grade GPU costs about 10-30
times less than the HPC CPUs and GPUs. But it's performance is not
reduced by the same factor. The consumer-grade GPU outperforms the
more expensive 20 core CPU on large grid-sizes and it's performance is
nearly on a par with the (much more expensive) 96 core CPU on the
largest possible grids.  This indicates that implementing a code using
Kokkos in order to target GPUs is interesting also from a financial
point of view.

\begin{figure}[t]
\centering
\includegraphics[width=0.48\textwidth]{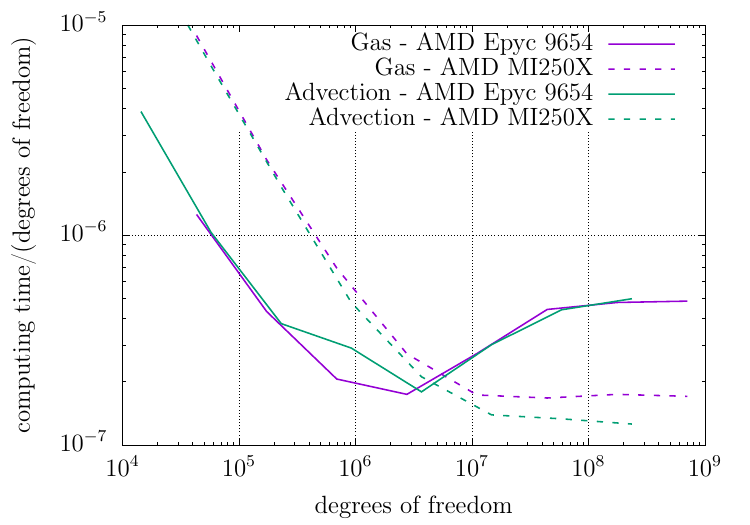}
\caption{\added{Computing time for performing 100 time steps of the
  two-dimensional advection (\ref{eq:advection2d}) and isothermal
  Euler equations (\ref{eq:eulerIso2d}) normalized to the number of
  degrees of freedom on a CPU and
  GPU. }}\label{fig:time_cell_d_eulerIso2d}
\end{figure}

\changed{In order to estimate the non-linear performance of the code
  we compare in Fig.~\ref{fig:time_cell_d_eulerIso2d} the performance
  of the advection (\ref{eq:advection2d}) and the isothermal Euler
  equations (\ref{eq:eulerIso2d}) and a CPU and GPU system. The curves
  for CPUs and GPUs fall nearly on top of each other. This is again
  showing that the overall number of degrees of freedom is a relevant
  quantity when comparing different computations. When comparing the
  computing time more in detail for the Euler- to the advection
  equation on has to keep in mind that the Euler equation requires
  more operations per degree of freedom for the flux calculations than
  the advection equation (see $\bF_x$ and $\bF_y$ in
  (\ref{eq:advection2d}) and (\ref{eq:eulerIso2d})). In this respect
  it is remarkable that the Euler equation only needs the same
  computing time as the advection equation on CPUs.  It is possible
  that the SIMD vectorization capacities, i.e. ability to process
  several variables at the same time, of the CPU is able to speed up
  vector-type calculations of the Euler equations.}

The present numerical code combines MPI communication and GPU
offloading so that simulations can be run on multiple GPUs. In the
recent past, data could only be transferred between different GPUs by
passing it through the hosting CPUs via MPI communication which
typically degraded the multi-GPU performance. Today's modern network
architectures allow to directly transfer data between GPUs. We note
that the data storing \verb+Kokkos::View+ is functionally identical to
a memory pointer so that \verb+Views+ can easily be used within MPI
functions. Depending on the amount of data to be shipped the direct
data transfer can considerably increase the multi GPU performance. For
the present numerical scheme, we measure a gain only of the order of
10\% to 20\% percent (not shown). This is because the scheme is very
local and only a relatively small amount of data needs to be exchanged
during the computations.

\begin{figure}[t]
\centering
\includegraphics[width=0.48\textwidth]{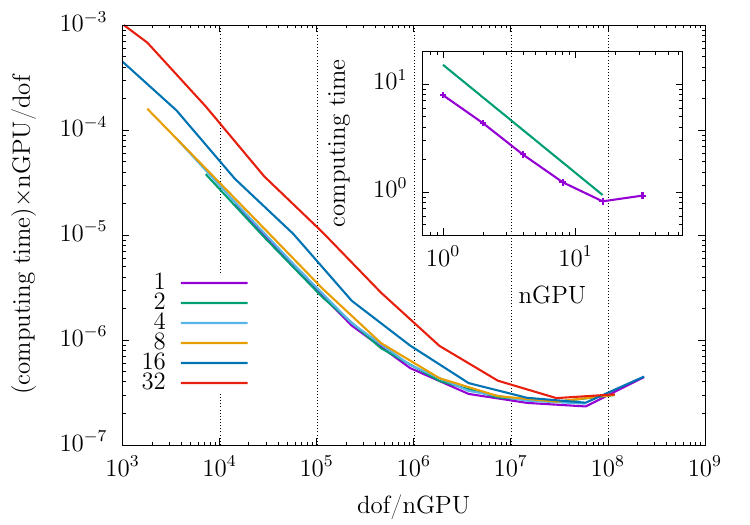}
\caption{\added{Computing time per degree of freedom on NVidia V100
    GPUs as a function of the number of degrees of freedom per
    GPU. Different curves belong to different number of GPUs ($nGPU$).
    The GPUs are connected via an Omni-Path interconnection network
    with 100 Gb/s. Inset: Computing time for roughly $6\times10^7$
    degrees of freedom as a function of the number of used
    GPUs}}\label{fig:time_cell_cuda_v70_jz}
\end{figure}

\begin{figure}[t]
\centering
\includegraphics[width=0.48\textwidth]{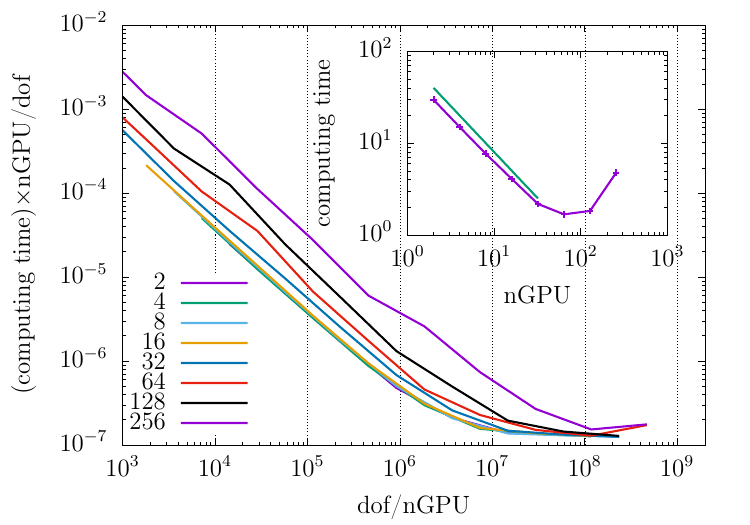}
\caption{\added{Computing time per degree of freedom on AMD MI250x
    GPUs as a function of the number of degrees of freedom per
    GPU. Different curves belong to different number of GPUs
    ($nGPU$). The GPUs are connected via a Slingshot 200 Gb/s Network
    Interface. Inset: Computing time for $2.36\times10^8$ degrees of
    freedom as a function of the number of used
    GPUs}}\label{fig:time_cell_rocm}
\end{figure}

We checked the multi-GPU performance of the RK-NDG scheme on two
different supercomputers using two different GPU architectures and
networks. In Fig.~\ref{fig:time_cell_cuda_v70_jz} we analyze the
speed-up of many NVidia V100 GPUs compared to a single one as a
function of the degrees of freedom treated per GPU. For perfect
scaling all curves should fall on top of each other. This is nearly
the case up to eight GPUs. \added{Weak scaling, i.e. keeping the
  number of degrees of freedom per GPU constant, can be read off along
  the vertical lines. For large grid sizes beyond $10^7$ dof weak
  scaling can be observed up to 16 GPUs. On smaller grids, the
  performance decreases already for 8 GPUs. This shows also in the
  context of multi-GPU computing the necessity of sufficient work-load
  in order to take full benefit of GPUs. The inset shows data on
  strong scaling, i.e. the computing time for a fixed (large) grid
  size. One observes quit good scaling up to 8 GPUs}. We note that
each compute node of the used supercomputer contains four GPUs. For
large grid-sizes, the performance on 32 GPUs is nevertheless still
good. The performance on smaller grid-sizes might suffer from
inter-GPU communication. Doing the same analysis on a different
supercomputer composed of AMD MI250X GPUs and a twice as fast network
the scaling is much better (see Fig.~\ref{fig:time_cell_rocm}). Here,
simulations on large grid-sizes perform very well even up to 64
GPUs. \added{Weak scaling can be observed for large grid sizes beyond
  $4\times10^7$ dof weak scaling can be observed up to 128 GPUs. On
  smaller grids, the performance decreases already for 32 GPUs. From
  the inset, one observes quite good strong scaling up to 32 GPUs on
  large grids.}

\begin{figure}[t]
\centering
\includegraphics[width=0.48\textwidth]{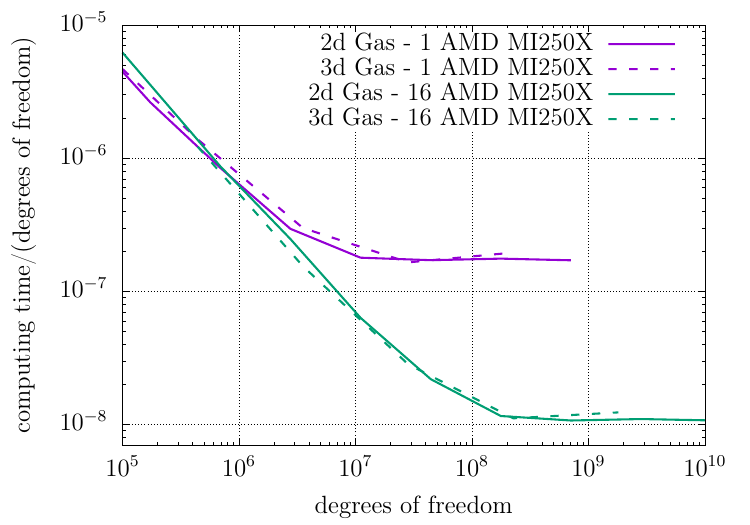}
\caption{\added{Computing time for performing 100 time steps of the
    two-dimensional and three-dimensional isothermal Euler equations
    (\ref{eq:eulerIso2d}) normalized to the number of degrees of
    freedom on a single GPU and on 16
    GPUs. }}\label{fig:time_cell_eulerIso2d_3d_d}
\end{figure}

\added{Let us finally show that a three-dimensional simulation shows
  that same performance as a two-dimensional simulation. In
  Fig.~\ref{fig:time_cell_eulerIso2d_3d_d} the computing time is shown
  as a function of the degrees of freedom of simulations of the
  isothermal Euler equations. The 2d and 3d curves fall on top of each
  for mono-GPU- as well as for multi-GPU simulations so that the 2d
  and 3d simulations perform equally well. }

Let us compare \changed{a multi-GPU simulation} to a possible multi-CPU
simulation. Because an AMD MI250X GPU is roughly four times faster
(see Fig.~\ref{fig:time_cell_d}) than the Intel 96 core CPU on large
($10^8$ degrees of freedom) grids, this implies that a 64 GPU
simulation, having close to perfect efficiency, corresponds to a 256
CPU simulation with 96 cores (24576 cores in total).

\subsection{Environmental considerations}
\label{subsec:env}
\begin{figure}[t]
\centering
\includegraphics[width=0.48\textwidth]{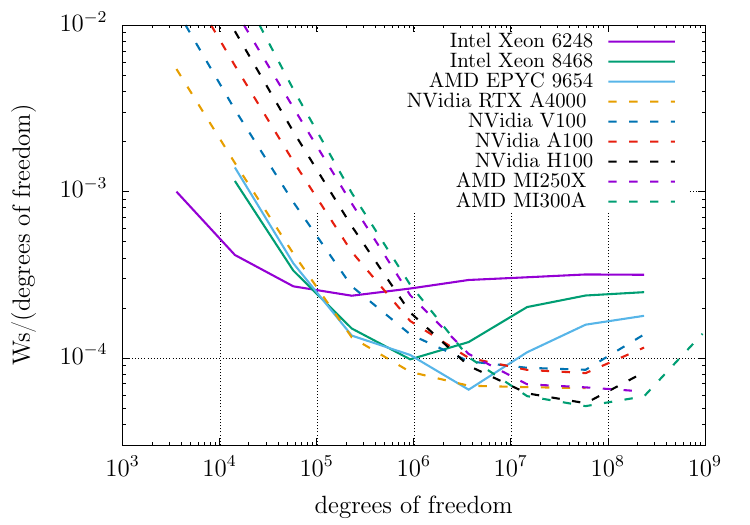}
\caption{Energy consumption for performing 100 time-steps of the
  two-dimensional linear advection equation (\ref{eq:advection2d}) per
  degree of freedom on various architectures.}\label{fig:ws_cell_d}
\end{figure}

The problem of climate change is great importance as certified in the
2016 Paris agreement of the United Nations~\cite{paris2016} and
high-performance computing should also take environmental
considerations seriously into account~\cite{european_hpc}. We shall
here estimate the environmental impact of numerical simulations using
the present numerical scheme. Today's supercomputer make heavy use of
GPU accelerators because of their higher energy efficiency compared to
CPUs. Indeed, the best ranked supercomputers in the Green 500 list
(see \cite{green500}) use GPUs. We have already seen that GPUs are
faster than CPUs for large grid-sizes while CPUs are faster on small
grid-sizes. However, the CPU table~\ref{table:cpu} and the GPU
table~\ref{table:gpu} document that GPUs typically consume more power
than CPUs. We therefore consider here the electric energy consumption
per simulated degrees of freedom estimated from the thermal design
power of the different devices.  Fig.~\ref{fig:ws_cell_d} demonstrates
that GPUs are indeed more energy efficient than CPUs on large
grids. Above $10^7$ degrees of freedom corresponding to $\approx
3100\times3100$ grids GPUs significantly outperform all considered
CPUs. However, below such high grid-sizes the most modern 96 core CPU
is more efficient than the HPC GPU hardware. Interestingly, the
efficiency of this CPU is very close to that of the consumer graphic
card. As the latter is also performing very well on large grids we
find that the consumer graphic card is also a good choice from an
energetic point of view. Finally, we note that on quite small grids
below $2\times 10^4$ degrees of freedom ($\approx 140 \times 140$) the
oldest 20 core CPU is the most efficient.

\begin{figure}[t]
\centering
\includegraphics[width=0.48\textwidth]{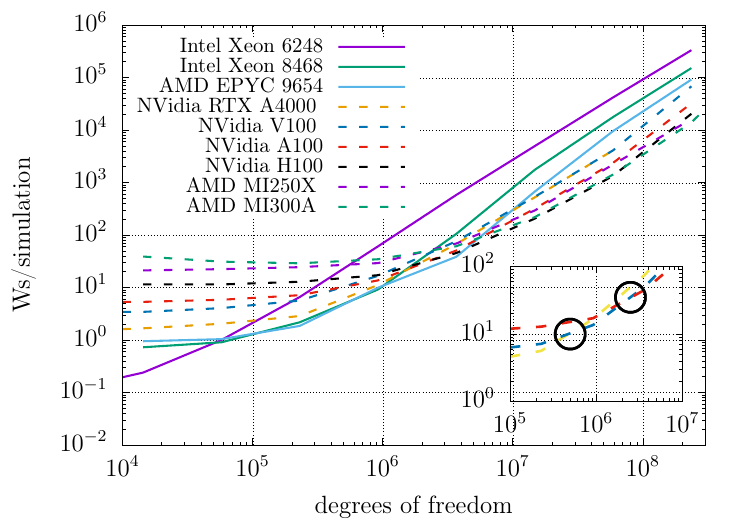}
\caption{Electric energy consumption for a 2d advection simulation
  performing one complete revolution. Inset: zoom onto the region
  where the three NVidia GPUs cross. The V100 and A100 cross at
  roughly $5\times 10^5$ grid points while the A100 and H100 cross at
  $2.5\times 10^6$.}\label{fig:ws_T_cell_d}
\end{figure}

Until now, we studied the energy consumption per degree of
freedom. Let us now turn to the energy needed to perform a complete
simulation. In order to estimate the total energy consumption of a
simulation we need to take into account that the required number of
time steps typically depends on the grid-size. In the case of the
simple two-dimensional advection equation (\ref{eq:advection2d})
stability constraints imply that the time step scales as
$\sqrt{1/dof}$ so that large-grid simulations need more time steps
than small-grid simulations. From Fig.~\ref{fig:ws_T_cell_d} we
observe that the total energy consumption of a simulation is strongly
increasing with the grid-size. It is worth to note that the higher
energetic efficiency of modern GPUs compared to CPUs is far from
compensating this increase of consumed energy. For example,
simulations with $10^8$ degrees of freedom that perform very well on
GPUs need roughly a thousand times more energy than $10^5$ degrees of
freedom simulations on CPUs.

As a final result we show that the more modern is the architecture the
larger needs to be the grid in order to compute efficiently. The
tendency is that new hardware is optimized for ever growing
simulations: By focusing on three different generation of HPC NVidia
GPUs (V100, A100, H100) we see in the inset of
Fig.~\ref{fig:ws_T_cell_d} that the more recent is the GPU the larger
needs to be the simulation in order to be more energy efficient than
the former architecture. While beyond $5\times 10^5$ degrees of
freedom the A100 consumes less energy than the V100, the H100 needs at
least $2.5\times 10^6$ degrees of freedom to beat the A100. The
minimal number of degrees of freedom for optimal usage of the compute
resources increases therefore by a factor of five from one generation
to the next. In conjunction with the former result that larger
simulations need more computing time than small simulations we
therefore observe a rebound effect~\cite{Biewendt-etal-2020}. This
means that the power efficiency gained by architectural improvements
are overshadowed by the increased energy consumption when switching to
larger grids that are appropriate to the new GPU generations.


\section{Conclusions and perspectives}
\label{sec:conc}

This paper analyzes the performance of a nodal discontinuous Galerkin
scheme (NDG) on modern HPC hardware using the Kokkos library in the
context of fluid dynamical simulations. We especially paid attention
to the performance gain obtained by using state of art GPUs compared
to CPUs. We find that NDG schemes are well suited for solving
demanding problems arising in fluid dynamics due to their high-order
of convergence and their locality. The former allows to solve a
problem with a given accuracy much faster than low-order schemes. It's
locality allows for a very efficient usage of modern massive parallel
architectures.

We show that an implementation of NDG type schemes using the Kokkos
library in conjunction with the Message Passing Interface (MPI) allows
for highly efficient simulations both on multi-CPU systems as well as
on multi-GPU systems. With a single code base and no noteworthy
architectural fine tuning we observe perfect weak-scaling with more
than 6000 CPU cores and 64 GPUs. We underline that the code runs
without modification on a variety of CPUs and GPUs of different
vendors. This shows that the NDG-Kokkos-MPI combination is valuable
for HPC demanding problems. Having only to maintain one single code
base is efficient from a programming perspective in this ever growing
complexity of HPC hardware.

We estimated the environmental impact of numerical simulations using
the NDG-Kokkos-MPI combination by estimating the energy consumption of
a simulation. We observe the well known fact that simulations on GPUs
consume less energy than on CPUs. But this is only true for large
scale simulations. Below a certain grid-size, CPUs are more
efficient. Furthermore, even when choosing the most adapted computing
architectures depending on the grid-size, the energy consumption of a
simulation depends roughly linearly on the number of grid points.  We
also observe a tendency that a new GPU generation only beats the old
one in terms of energy efficiency when switching to ever larger
grids. But the efficiency gain of new GPUs does not overcome the
increase of the energy consumption for these larger simulations.

In a future publication we plan to extent this work to different fluid
equations and to investigate performance gains from fine-tuning of the
code. We further plan to directly compare the performance of the
NDG-Kokkos-MPI combination to a pseudo-spectral code. Pseudo-spectral
codes are widely used in the HPC fluid dynamics community because they
offer the highest order of convergence. Their drawback is the
mandatory use of Fourier transformations in conjunction with massive
global data transfer. It will be therefore interesting to compare
their performance on modern HPC GPU hardware to the NDG-Kokkos-MPI
code discussed in this paper.

\section*{Acknowledgments}
This work was granted access to the HPC resources of IDRIS under the
allocation 2024-AD012A15700 and to the HPC resources of CINES under
the allocation 2024-AD012A15823 made by GENCI.

This work was supported by the French government through the France
2030 investment plan managed by the National Research Agency (ANR), as
part of the Initiative of Excellence Université Côte d’Azur under
reference number ANR-15-IDEX-01. The authors are grateful to the
Université Côte d’Azur’s Center for High-Performance Computing (OPAL
infrastructure) for providing resources and support.

Part of the project was conducted in preparation for a Master's thesis
for MAUCA, Master track in Astrophysics in Université Côte
d'Azur. This Master's thesis was funded by EUR Spectrum of the
Université Côte d'Azur.






  
\end{document}